\documentclass[11pt,a4paper,oneside]{article}
\usepackage{mathptmx}

\begin{document}

\title{Semiclassical study on Proton and Neutron}

\author{
	I. V. Grossu(1), C. Besliu(1), Al. Jipa(1), D. Felea(2) and C. C. Bordeianu(1)\\\\
	(1) University of Bucharest, Faculty of Physics, Bucharest-Magurele, Romania\\
	(2) Institute of Space Sciences, Bucharest-Magurele, Romania\\
}

\maketitle

\begin{abstract}
Starting from the existing semiclassical studies on hydrogenoid atoms, we propose a similar intuitive exercise for the three-body quark systems corresponding to protons and neutrons. In the frame of this toy model we try to explain both the stabilities of proton and neutron with respect to the nuclear interaction, and the spectrum of nucleonic resonances with $J=1/2$. Our choice is motivated also by a good agreement obtained for the up and down quark rest masses report. Taking into account the deterministic chaotic behavior of many-body systems, the discussed exercise could be understood as an interesting particular case of a quantum three-body problem which admits a semiclassical treatment.
\end{abstract}

\section{Introduction}

As the evolution of many-body systems \cite{Bib:Nbody} is, generally, very complex, a semiclassical treatment encounters important difficulties. Inspired by existing nuclear billiard models \cite{Bib:Billiard01,Bib:Billiard02}, we tried to apply chaos theory for the three-body quark systems corresponding to nucleons. In this context we noticed that, for the particular exactly solvable case of harmonic potential \cite{Bib:Harmonic}, the closing trajectories condition can be written in function of the constituent masses \cite{Bib:Conf} as a simple Lissajous equation. Based on this observation, one can consider stationary de Broglie waves \cite{Bib:Broglie} for the quark components of proton and neutron.

\section{Study on Proton and Neutron}

One first problem involved by this approach is related to the quark-quark interaction expression. In agreement with the Isgur-Karl quark model \cite{Bib:Isgur}, we assume the harmonic potential plays a significant role in this context. Another justification could be the fact that, by developing the potential in Taylor series, the harmonic term could be a good choice according to the asymptotic freedom property. 

We started from an even simpler case of the three-body problem with harmonic interactions, as two of the constituents have the same mass (up, up, down, respectively down, down, up, quark structures). By considering the particles are initially at rest, placed in the vertices of an arbitrary triangle, the motion could be decomposed with respect to the geometric center of the system. Thus, the movement of all constituents contains a harmonic component with angular frequency $\omega_1$. The motions of the two particles with the same mass are similar, and can be obtained by superposing an additional non-parallel harmonic oscillation with frequency $\omega_2$. In this context, the existence of closed trajectories \cite{Bib:Conf} is assured by simply considering a Lissajous condition:

\begin{equation}
\label{Eq:Lissajous} 
\frac{{\rm \omega }_{{\rm 2}} }{{\rm \omega }_{{\rm 1}} } =\sqrt{\frac{3m_{2} }{2m_{1} + m_{2} } }  {\in }  {{\it Q}} 
\end{equation}
where $m_1=m_3$, $m_2$ are the particle masses, $\omega_1$ is the frequency common to all the three constituents, $\omega_2$ the frequency of the non-parallel harmonic oscillator, and $Q$ is the set of rational numbers.

In agreement with the known stabilities of protons and neutrons, we tried to find those values for the up and down quark masses report $r=m_u/m_d$, which satisfy the Eq. (\ref{Eq:Lissajous}) for both (up, up, down), and (down, down, up) quark structures. As no solution was found, a less restrictive condition was imposed. We thus searched for a report which satisfies the closing trajectories condition for proton, and only approximates it for neutron. Among the various results (e.g. $r=11/32, r=23/75$ etc.) we noticed the
value $r=13/24=0.541(6)$, which is very close to the known theoretical value $0.56$ \cite{Bib:Reportupdown}.

Another problem is related to the possibility of directly applying stationary de Broglie waves \cite{Bib:Broglie}, as the analyzed three-body system is more complex than, for example, the case of hydrogenoid atoms \cite{Bib:Bohr}.  One possible workaround solution is based on using the well known energy spectrum of the one-dimensional quantum harmonic oscillator:

\begin{equation}
\label{Eq:Qharmonic} 
E_{n} =\left(n+\frac{1}{2} \right)\hbar \omega  
\end{equation} 
where $n$ is an integer $\geq0$, $\hbar$ the Plank constant, and $\omega$ the frequency.

The relativistic effects cannot be neglected also. One improvement could come from using the spectrum of the Dirac oscillator \cite{Bib:Dirac} which, for $n<<mc^2/2h\omega$, is approximated with Eq. (\ref{Eq:Qharmonic})).
 
As previously mentioned, for the discussed particular 3-body problem, the motion of all the three constituents contains a harmonic component with frequency $\omega_1$. For the two particles with the same mass, we must take into account also an additional non-parallel harmonic oscillation with frequency $\omega_2$. One can consider the system as a collection of five harmonic oscillators (three with frequency $\omega_1$, and two with $\omega_2$). In ground state, the total energy must equals the rest mass of proton. The following system of equations is obtained:

\begin{equation} 
\label{Eq:Psys} 
\left\{\begin{array}{c} {\frac{\hbar }{{\rm 2}} \left(3{\rm \omega }_{{\rm 1p}} +2{\rm \omega }_{{\rm 2p}} \right)=m_{p} c^{2} } \\ {\frac{{\rm \omega }_{{\rm 2p}} }{{\rm \omega }_{{\rm 1p}} } =\sqrt{\frac{3m_{d} }{2m_{u} +m_{d} } } } \end{array}\right.  
\end{equation} 
where $m_p$ is the mass of proton, and $m_u$, $m_d$  represent the masses for the up and down quarks. 

Using the earlier specified value $r=m_u/m_d=13/24\simeq0.56$, and working in the natural system of unities, we obtained $\omega_{1p}\simeq347MeV$, and $\omega_{2p}\simeq417MeV$.

The analog system for neutron is:

\begin{equation} 
\label{Eq:Nsys} 
\left\{\begin{array}{c} {\frac{\hbar }{{\rm 2}} \left(3{\rm \omega }_{{\rm 1n}} +2{\rm \omega }_{{\rm 2n}} \right)=m_{n} c^{2} } \\ {\frac{{\rm \omega }_{{\rm 2n}} }{{\rm \omega }_{{\rm 1n}} } =\sqrt{\frac{3m_{u} }{2m_{d} +m_{u} } } } \end{array}\right.  
\end{equation} 
where $m_n$ is the mass of neutron.

We obtained $\omega_{1n}\simeq408MeV$, and $\omega_{2n}\simeq326MeV$. It is interesting to notice that $\omega_{1p}\simeq\omega_{2n}$ and $\omega_{2p}\simeq\omega_{1n}$.

Based on this simplified study in which, for example, the spin and the electric charge were ignored, we calculated the excited states of nucleons. Despite all approximations, one can notice an encouraging agreement (Table 1) obtained for the spectrum of nucleonic resonances with $J=1/2$ (with one exception, all computed  values differ from the known corresponding masses \cite{Bib:Pdg} with less than 6.5\%). The missing resonance could be understood as a problem of the
discussed model. One can predict also the mass for a possible new nucleonic resonance with $J=1/2$.

\begin{table}[t]
\begin{tabular}{ccccccc}
\hline 
Resonance & \multicolumn{3}{p{1.0in}}{Proton} & \multicolumn{3}{p{1.0in}}{Neutron} \\ \hline 
 Mass\newline (MeV) \cite{Bib:Pdg} & Level\newline  & Mass\newline (MeV) & $\Delta m/m$ & Level & Mass\newline (MeV) & $\Delta m/m$ \\ \hline 
- & $\omega $1 & 1285 & - & $\omega $2 & 1266 & - \\ \hline 
1440  & $\omega $2 & 1355 & 6\% & $\omega $1 & 1348 & 6.4\% \\ \hline 
1535  & 2$\omega $1 & 1633 & 6.4\% & 2 $\omega $2 & 1592 & 3.7\% \\ \hline 
1650 & $\omega $1+ $\omega $2 & 1702 & 3.1\% & $\omega $1+ $\omega $2 & 1674 & 1.5\% \\ \hline 
1710 & 2 $\omega $2 & 1772 & 3.6\% & 2$\omega $1 & 1756 & 2.7\% \\ \hline 
2090 & 2$\omega $1+ $\omega $2 & 2049 & 2\% & 2$\omega $2+ $\omega $1 & 1999 & 4.3\% \\ \hline 
2100 & 2$\omega $2+ $\omega $1 & 2119 & 1\% & 2$\omega $1+ $\omega $2 & 2081 & 1\% \\ \hline 
New? & 2$\omega $2+ 2$\omega $1 & 2467 & ? & 2$\omega $1+ 2$\omega $2 & 2407 & ? \\ \hline 
\end{tabular}
\caption{The excited states of nucleons, calculated using Eq. (\ref{Eq:Qharmonic},\ref{Eq:Psys},\ref{Eq:Nsys}), compared with the spectrum of nucleonic resonances with $J=1/2$}
\label{Tab:Resonances}
\end{table}

\section{Conclusions}

The chaotic behavior \cite{Bib:Nbody,Bib:Billiard02} of three-body systems is incompatible with a semiclassical treatment. However, for the particular case of harmonic potential \cite{Bib:Harmonic}, the closing condition of the constituent trajectories becomes a simple Lissajous equation Eq.(\ref{Eq:Lissajous}). Based on this observation and inspired by existing studies on hydrogenoid atoms \cite{Bib:Bohr}, we propose an intuitive semiclassical analysis for the more complex case of the three-body problem with harmonic interactions.  In the frame of this simplified toy model we tried to explain the proton and neutron stabilities with respect to the nuclear interaction. We obtained a good value for the up and down quark masses report $r=m_u/m_d=13/24\simeq0.56$ \cite{Bib:Reportupdown}, and an encouraging agreement for the spectrum of nucleonic resonances with $J=1/2$ (Table 1). In this context, the discussed exercise could be understood as an interesting example of a particular quantum three-body problem which admits a semiclassical treatment.

\end{document}